\documentclass[aps,prl,reprint,letterpaper,nolongbibliography,nofootinbib]{revtex4-2}
\usepackage[T1]{fontenc}
\usepackage{amsmath,amssymb,bm,graphicx,booktabs,microtype}
\usepackage{comment}
\usepackage[dvipsnames]{xcolor}
\usepackage[colorlinks=true,linkcolor=MidnightBlue,citecolor=MidnightBlue,urlcolor=MidnightBlue]{hyperref}
\newcommand{\ket}[1]{\lvert #1\rangle}
\newcommand{\bra}[1]{\langle #1\rvert}
\newcommand{\SM}{\mathrm{SM}}

\newcommand{\rs}{r_s}
\newcommand{\GeV}{\,\mathrm{GeV}}
\newcommand{\TeV}{\,\mathrm{TeV}}
\newcommand{\Mtwo}{\langle\mathcal{M}_2\rangle}

\begin{document}
\raggedbottom
\title{Quantum Magic  and the Strong Coupling Constant}

\author{ Qiaofeng Liu$^{\, a}$, Ian Low$^{\, a,b}$, Zhewei Yin$^{\, c}$}

\affiliation{\vspace{0.1cm}
\mbox{$^a$ Department of Physics and Astronomy, Northwestern University, Evanston, IL 60208, USA}\\ 
\mbox{$^b$ High Energy Physics Division, Argonne National Laboratory, Lemont, IL 60439, USA}\\
\mbox{$^c$ Institute of High Energy Physics, Chinese Academy of Sciences, Beijing 100049, China}\\
 \vspace{-0.3cm}}
 
\begin{abstract}
The three gauge couplings of the Standard Model (SM)  can be parameterized as the fine structure constant $\alpha$, the weak mixing angle $\sin^2\theta_w$, and the strong coupling $\alpha_s$. They are fundamental constants whose values remain unexplained. Previously we showed that minimizing  magic production in charged-lepton scattering reproduces $\sin^2\theta_w$ with great precision. Here we study magic production in the six flavor-diagonal color-singlet channels $q\bar q\to q\bar q$ at tree level, including photon, gluon, $W/Z$, and Higgs exchange. Varying only $\alpha_s$, with all other inputs fixed at each center-of-mass energy, we find that the finite-$\alpha_s$ magic minimum in the top channel reproduces the $\overline{\mathrm{MS}}$ value of $\alpha_s$ to within $5\%$ from  $1$~TeV to $10^9$~GeV. Higgs exchange involving the large top Yukawa coupling is essential for this agreement. All six channels produce near-minimal magic at the SM couplings, and reducing $\alpha_s$ toward $\alpha$ increases their magic production, potentially explaining why the strong interaction is ``strong.''  To understand why minimizing magic reproduces both $\sin^2\theta_w$ and $\alpha_s$, we conjecture that the parameters of fundamental interactions among color singlets reflect a principle of quantum computational efficiency, with the physical universe emerging from a quantum simulation subject to constrained resources.
\end{abstract}
\maketitle

\section{Introduction}
\label{sec:intro}
The gauge interactions of the Standard Model (SM) are determined by the symmetry group $SU(3)_c\times SU(2)_L\times U(1)_Y$ with coupling constants $g_3$, $g_2$ and $g_1$, respectively. The coupling strengths define three of the most important fundamental constants in nature: the electric charge $e\equiv g_1g_2/\sqrt{g_1^2+g_2^2}$, the weak mixing angle $\sin^2\theta_w \equiv g_1^2/(g_1^2+g_2^2)$ and the strong coupling $g_s\equiv g_3$.  Their values are unexplained and precisely measured experimentally at several energy scales~\cite{ParticleDataGroup:2024cfk}. 

In Ref.~\cite{Liu:2025bgw} we initiated a study of whether the observed couplings are special from the perspective of quantum computation. The relevant resource is non-stabilizerness, or magic, introduced by Bravyi and Kitaev~\cite{Bravyi:2004isx}. By the Gottesman--Knill theorem~\cite{Gottesman:1998hu,Gottesman:1999tea,Aaronson:2004xuh}, circuits built from Clifford gates acting on stabilizer states can be simulated efficiently on a classical computer even when they generate maximal entanglement, so entanglement alone is insufficient for a quantum speedup, motivating magic as an additional computational resource. In Ref.~\cite{Liu:2025bgw}, minimizing the production of magic in charged-lepton scattering gave a weak mixing angle within a percent of its measured $\overline{\rm MS}$ value at around the TeV scale. Because an overall factor of the scattering amplitude drops out of the normalized outgoing spin state at tree level and at fixed center-of-mass (CM) energy,  the magic production in this case  depends on the gauge couplings only through $\sin^2\theta_w$.

The quantum-information perspective on fundamental interactions has by now a substantial literature. The emergence of symmetries from the suppression or maximization of entanglement or magic in scattering has been studied in a variety of settings~\cite{Beane:2018oxh,Low:2021ufv,Liu:2022grf,Carena:2023vjc,Liu:2023bnr,Hu:2024hex,Chang:2024wrx,Kowalska:2024kbs,McGinnis:2025brt,Busoni:2025dns,Carena:2025wyh,Hu:2025lua,McGinnis:2025iab,McGinnis:2025xgt,Li:2026kha,Low:2026evp,Li:2026udy,Kanemura:2026won,Low:2026oyf, Low:2026kxb}. An area law relates the entanglement entropy to the elastic cross section~\cite{Aoude:2024xpx,Low:2024mrk,Low:2024hvn,Liu:2025pny}. Entanglement has also been related to SM parameters, as in $\sin^2\theta_w=1/4$ from maximal entanglement~\cite{Cervera-Lierta:2017tdt}, flavor mixing from entanglement minimization~\cite{Thaler:2024anb}, and parameter inference in Higgs decays~\cite{Liu:2025iwh}. Magic production has been computed for processes in nuclear and particle physics~\cite{Robin:2024oqc,Li:2026udy,Chernyshev:2024pqy,Liu:2025qfl,White:2024nuc,Aoude:2025jzc,Gargalionis:2025iqs,Gargalionis:2026onv,Chu:2026yxm,Alvarado:2026owd,Viaux:2026skf,Cao:2026aye}, alongside studies of maximal and nonlocal magic~\cite{Liu:2025frx,Roman:2026mcy,Knipfer:2026ixm,Robin:2025dgw,Busoni:2026lvp}. Notably, the magic produced depends on the spin basis, since the rotation from laboratory to helicity bases is generally not a Clifford operation, but the physical weak mixing angle lies close to the magic-minimizing value in both bases~\cite{Li:2026zum}. These results motivate computing the magic production in quark scatterings,  where the strong, electroweak, and Yukawa interactions all contribute.

\begin{figure*}[t]
 \centering
 \includegraphics[width=0.8\textwidth]{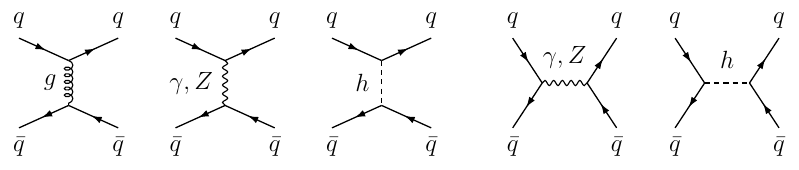}
 \caption{Tree diagrams for flavor-diagonal color-singlet $q\bar q\to q\bar q$ scattering. Notice that the $s$-channel gluon exchange vanishes in the color singlet channel.}
 \label{fig:diagrams}
\end{figure*}

In this work we study elastic quark scattering at tree level, focusing on the six flavor-diagonal color singlets $q\bar q\to q\bar q$ (Fig.~\ref{fig:diagrams}). Gluon exchange introduces an additional gauge-coupling ratio, $\alpha_s/\alpha$, in the normalized spin state, beyond the weak mixing angle probed in charged-lepton scattering. Following Ref.~\cite{Liu:2025bgw}, we use the laboratory (Lab) basis and vary only $\alpha_s$, holding all other inputs fixed. Near the TeV scale $\alpha_s/\alpha\simeq11$, and the normalized spin states of the light-quark channels are close to those produced by gluon exchange alone, so that their magic depends only weakly on $\alpha_s$ near the SM value. Near-minimal magic in these channels therefore does not by itself determine $\alpha_s$ precisely. In the top channel, however, the large top Yukawa coupling allows Higgs exchange to compete with gluon exchange. Physical Higgs exchange is essential for a deep, finite-$\alpha_s$ magic minimum whose location reproduces the $\overline{\rm MS}$ running $\alpha_s$ to within a few percent from about $1\TeV$ to $10^8\GeV$. At higher CM energies gluon exchange becomes less dominant, with $\alpha_s/\alpha\simeq2.5$ at $10^{16}\GeV$, yet all six diagonal channels remain close to their magic minima. Moreover, reducing $\alpha_s$ to $\alpha$ increases their magic production, offering a potential explanation for why the strong interaction is stronger than electromagnetism. Flavor-nondiagonal and nonsinglet channels show a less uniform pattern.

\section{Stabilizer R\'enyi entropy}
\label{sec:magic}
Characterizing and quantifying magic is an active subject in quantum computation~\cite{Veitch:2012ttw,Mari:2012ypq,Veitch:2013bll,Bravyi:2018ugg,Liu:2019qfi,Liu:2020yso,Warmuz:2024cft}. Here we briefly review the measure used below, the stabilizer R\'enyi entropy (SRE)~\cite{Leone:2021rzd}, which is convenient to compute, can be measured experimentally~\cite{Oliviero:2022bqm,Haug:2023ffp}, and, for integer order $\alpha\ge2$, is a monotone under deterministic pure-state stabilizer protocols~\cite{Veitch:2013bll,Chitambar:2018rnj,Leone:2024lfr,Bittel:2025yhq}. For a normalized pure state of two qubits, the second SRE is
\begin{equation}
 M_2(\ket{\psi})=-\ln\!\left[\frac14\sum_{P\in\{I,X,Y,Z\}^{\otimes2}}
 \bigl(\bra{\psi}P\ket{\psi}\bigr)^4\right].
 \label{eq:magic}
\end{equation}
Here $P=\sigma_i\otimes\sigma_j$ runs over the sixteen two-qubit Pauli strings, where $\sigma_i\in\{I,X,Y,Z\}$ are the identity and the three Pauli matrices; see Refs.~\cite{Liu:2025qfl,Liu:2025frx} for brief reviews. Equation~\eqref{eq:magic} vanishes on stabilizer states and is invariant under Clifford transformations. We use the natural logarithm; the maximum of $M_2$ over two-qubit states is $\ln(16/7)\simeq0.827$~\cite{Liu:2025frx}. There are sixty pure two-qubit stabilizer states: thirty-six product states and twenty-four maximally entangled states. Their explicit construction is given in Appendix~A of Ref.~\cite{Li:2026zum}. Averaging the outgoing magic over this set provides a common comparison of scattering amplitudes whose inputs have no magic.

\section{Gauge couplings and running}
\label{sec:running}
With $e$, $\sin^2\theta_w\equiv s_w^2$, and $g_s$ defined in the beginning of Sec.~\ref{sec:intro}, the variable we scan is
\begin{equation}
 \rs\equiv\frac{g_s^2}{e^2}=\frac{\alpha_s}{\alpha}.
 \label{eq:couplings}
\end{equation}
We use the full-SM gauge and Yukawa couplings in the $\overline{\mathrm{MS}}$ scheme at the renormalization scale $\mu=\sqrt{s}$, the CM energy. These running couplings, together with the mass prescription specified below, define our renormalization-group-improved tree-level calculation. The one-loop evolution of the gauge couplings is
\begin{equation}
 \mu \frac{dg_i}{d\mu}=b_i \frac{g_i^3}{16\pi^2},\quad
 (b_1,b_2,b_3)=\left(\frac{41}{6},-\frac{19}{6},-7\right),
 \label{eq:running}
\end{equation}
where $g_1$ is the hypercharge coupling in its conventional normalization, without assuming grand unification~\cite{ParticleDataGroup:2026mpi}. We start the running at $\mu_0=91.188\GeV$, chosen equal to the input $Z$-boson mass. Equation~\eqref{eq:running} treats the full SM, including the top quark, as active from $\mu_0$, so its initial conditions are the six-flavor $\overline{\rm MS}$ couplings. We take $g_1$, $g_2$, and $g_3$ at $\mu_0$ from Ref.~\cite{Antusch:2025fpm}, where they are obtained with SMDR~\cite{Martin:2019lqd} from PDG~2024 inputs~\cite{ParticleDataGroup:2024cfk}; Table~\ref{tab:inputs} lists these and all other inputs. They correspond to $\alpha^{-1}(\mu_0)=128.11$, $s_w^2(\mu_0)=0.2315$, and $\alpha_s(\mu_0)=0.1170$, the last being the six-flavor counterpart of the top-decoupled value $\alpha_s^{(5)}(\mu_0)=0.1180$ used by the Particle Data Group (PDG).

\begin{table}[t]
\caption{Initial inputs at the starting scale $\mu_0=91.188\GeV$ and fixed mass inputs. The gauge and Yukawa couplings are full-SM $\overline{\rm MS}$ values from Ref.~\cite{Antusch:2025fpm}; the CKM parameters $s_{13}$, $s_{23}$, and $\delta$ are taken from the same reference. The starting scale is chosen equal to the input $Z$-boson mass, and this input, $M_W^{\mathrm{input}}$, $m_h$, and $|V_{ud}|$ are from Ref.~\cite{ParticleDataGroup:2024cfk}. The normalization $v$ and the Higgs propagator mass $m_h$ are held fixed.}
\label{tab:inputs}
\centering\setlength{\tabcolsep}{5pt}
\begin{tabular}{ll}
\toprule
$\mu_0$ & $91.188\GeV$\\
$g_1,\ g_2,\ g_3$ & $0.35727,\ 0.65096,\ 1.2123$\\
$y_u,\ y_c,\ y_t$ & $7.04\times10^{-6},\ 3.56\times10^{-3},\ 0.967$\\
$y_d,\ y_s,\ y_b$ & $1.54\times10^{-5},\ 3.06\times10^{-4},\ 0.01630$\\
$y_e,\ y_\mu,\ y_\tau$ & $2.78\times10^{-6},\ 5.85\times10^{-4},\ 0.00994$\\
$s_{13},\ s_{23},\ \delta$ & $0.00370,\ 0.04193,\ 1.139$\\
$|V_{ud}|$ & $0.97367$\\
$M_W^{\mathrm{input}}$ & $80.3692\GeV$\\
$v=2M_W^{\mathrm{input}}/g_2(\mu_0)$ & $246.925\GeV$\\
$m_h$ & $125.20\GeV$\\
\bottomrule
\end{tabular}
\end{table}

The three-generation Yukawa matrices are evolved with the full one-loop SM running in the $\overline{\mathrm{MS}}$ scheme~\cite{Antusch:2005gp,Kielanowski:2008wm}. We use the convention $Y_f=U_{Lf}\,\mathrm{diag}(y_f)\,U_{Rf}^{\dagger}$, so that the CKM matrix is $V=U_{Lu}^{\dagger}U_{Ld}$. The Yukawa couplings at $\mu_0$ are also taken from Ref.~\cite{Antusch:2025fpm}. The CKM matrix at $\mu_0$ is written in the standard unitary parametrization~\cite{Chau:1984fp}, with $s_{13}$, $s_{23}$, and $\delta$ from Ref.~\cite{Antusch:2025fpm} and $c_{12}=|V_{ud}|/c_{13}$ fixed by the measured $|V_{ud}|$ (Table~\ref{tab:inputs}). This choice defines a unitary benchmark combining the two input sources and gives $|V_{us}|=0.22793$. We define the running fermion and vector masses by
\begin{equation}
 \begin{gathered}
 m_f(\mu)=\frac{y_f(\mu)v}{\sqrt2},\qquad M_W(\mu)=\frac{g_2(\mu)v}{2},\\
 M_Z(\mu)=\frac{v}{2}\sqrt{g_2^2(\mu)+g_1^2(\mu)},
 \end{gathered}
 \label{eq:masses}
\end{equation}
with a fixed electroweak normalization $v=246.925\GeV$. We choose $v=2M_W^{\mathrm{input}}/g_2(\mu_0)$ so that the running $W$ mass reproduces the measured input at $\mu=\mu_0$. The $Z$-boson mass entering the amplitudes follows Eq.~\eqref{eq:masses}, giving $M_Z(\mu_0)=91.678\GeV$. The common $v$ in Eq.~\eqref{eq:masses} preserves the tree-level mass--coupling relations required for a consistent treatment of longitudinal $W/Z$ exchange and Yukawa interactions. Defining running fermion masses through running Yukawa couplings and a fixed electroweak normalization is an established convention~\cite{Fusaoka:1998vc,Xing:2007fb,Antusch:2013jca}, where the scale dependence of the masses in Eq.~\eqref{eq:masses}  comes entirely from the Yukawa and gauge couplings. These running masses are used consistently in external spinors, kinematics, and propagators.

Unless stated otherwise, we use the physical Higgs mass, $m_h=125.20\GeV$, as a fixed propagator input. The Higgs quartic coupling does not enter these tree amplitudes independently and does not appear in the one-loop evolution of the gauge and Yukawa couplings. Thus the running scalar potential is not needed for the prescription adopted here. The Higgs mass enters only through the propagator denominators.  All scattering amplitudes are evaluated at tree level with zero widths.

At each $\sqrt{s}$ we first evaluate the SM reference values of the running couplings, the CKM matrix, and the running masses of Eq.~\eqref{eq:masses} at $\mu=\sqrt{s}$. We then vary only $\alpha_s$, or equivalently $\rs$ since $\alpha$ is fixed. All other inputs, including $y_t$, the CKM matrix, and the masses, remain fixed throughout this scan. The SM point is $\rs^{\SM}=\alpha_s(\mu)/\alpha(\mu)$.

We display the coupling curves through $\rs=100$ only to show their approach to the pure-QCD limit, $\rs\to\infty$, which is equivalent to turning off all other couplings. We shade the region $\alpha_s>0.25$ as an illustrative benchmark for larger couplings, where higher-order corrections may become sizable. The shaded region begins at
\begin{equation}
 \rs^{(0.25)}=\frac{0.25}{\alpha(\mu)}.
 \label{eq:shade}
\end{equation}
The size of loop corrections  depends on the observable, kinematics, and renormalization scheme; this benchmark only serves as a visual reminder in the figures. The minimum of the averaged magic, $\Mtwo_{\min}(\sqrt{s})$ defined in Eq.~\eqref{eq:comparison}, is taken over the entire range $\rs\ge0$, including the asymptotic limit $\rs\to\infty$; neither the shading nor the displayed range truncates this minimization. In the asymptotic limit, the overall strong-coupling factor cancels from the normalized spin state. The resulting magic then arises purely from QCD, without the electroweak and Yukawa interactions.

\section{Magic in quark scattering}
\label{sec:scattering}

\begin{figure}[t]
 \centering
 \includegraphics[width=\columnwidth]{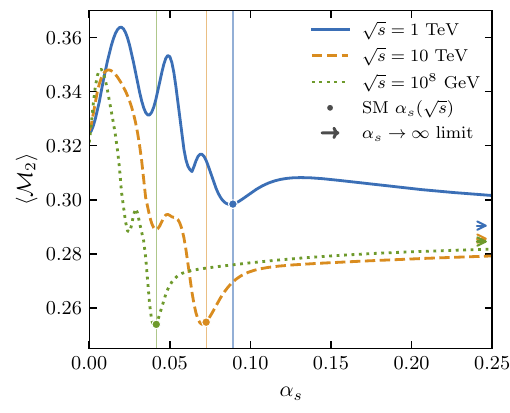}
 \caption{Magic production  in the $t\bar t$ color singlet as a function of $\alpha_s$ at $\sqrt{s}=1\TeV$, $10\TeV$, and $10^8\GeV$, with all other inputs fixed according to Sec.~\ref{sec:running}. Circles and thin vertical lines mark the SM value $\alpha_s(\sqrt{s})$. Arrows at the right edge mark the pure-QCD limit $\alpha_s\to\infty$.}
 \label{fig:ttcurves}
\end{figure}

Our choice of reference frame for spin projection is the Lab basis defined in Ref.~\cite{Liu:2025qfl}, where $\hat z$ is along the incoming quark, $\hat x$ lies in the scattering plane, and $\hat y$ is normal to it, with the axes oriented so that the outgoing quark has azimuthal angle $\phi=0$. In the Lab basis amplitudes and magic then depend on $\theta$ only. For both incoming and outgoing particles, the computational basis of qubits is defined by the spin projection onto the common $\hat z$ axis.

For each diagonal flavor we prepare and project the coherent color singlet $\ket{1}=\sum_{i=1}^3\ket{i\bar i}/\sqrt3$. The resulting spin amplitude $A(c_\theta;\rs,\sqrt{s})$, with $c_\theta\equiv\cos\theta$, includes all interfering diagrams in Fig.~\ref{fig:diagrams}. In the  singlet channel the color factors are $C_F=4/3$ for $t$-channel gluon exchange, and $1$ and $3$ for the exchange of a colorless boson in the  $t$-channel and  $s$-channel, respectively. The $s$-channel gluon exchange vanishes in the  singlet channel,  since a color-singlet $q\bar q$ pair cannot annihilate into a color-octet gluon. The dependence on $\alpha_s/\alpha$ thus comes entirely from the $t$-channel gluon term. The massive $W/Z$ contributions are evaluated in the unitary gauge.

For an initial stabilizer state $\ket{\psi_s}_i$, $i=1,\dots,60$, the spin configuration of the outgoing quarks at the scattering angle $\theta$ is given by the color-singlet spin amplitude and represented by the state vector $\ket{\tilde\psi}=A\ket{\psi_s}_i$. Since fixing the angle $\theta$ amounts to a projective measurement of the outgoing momentum direction, $\ket{\tilde\psi}$ is not normalized, and we compute the magic from $\ket{\psi}=\ket{\tilde\psi}/\langle\tilde\psi|\tilde\psi\rangle^{1/2}$, denoted $M_2(\ket{\psi_s}_i,\theta)$.
 Following Ref.~\cite{Liu:2025bgw}, we average over the sixty stabilizer states and the solid angle,
\begin{align}
 M_2(\theta)&\equiv\frac1{60}\sum_{i=1}^{60}M_2(\ket{\psi_s}_i,\theta),\notag\\
 \Mtwo(\rs;\sqrt{s})&\equiv\frac1{4\pi}\int M_2(\theta)\,\sin\theta\,d\theta\,d\phi,
 \label{eq:average}
\end{align}
where the dependence of $M_2(\theta)$ on $\rs$ and $\sqrt{s}$ is left implicit. The angular average is uniform in solid angle and is not weighted by the differential cross section. 

We proceed to minimize Eq.~\eqref{eq:average} with respect to $r_s$ over the full coupling range, including its asymptotic limit. At each CM energy the minimum and percentage excess at the SM point are
\begin{align}
 \Mtwo_{\min}(\sqrt{s})&\equiv\inf_{\rs\ge0}\Mtwo(\rs;\sqrt{s}),\notag\\
 \delta_{\Mtwo}(\sqrt{s})&\equiv100\,\frac{\Mtwo(\rs^{\SM};\sqrt{s})-\Mtwo_{\min}(\sqrt{s})}{\Mtwo_{\min}(\sqrt{s})}.
 \label{eq:comparison}
\end{align}
In the following we begin with the top channel, where the Higgs exchange contribution is comparable to that from the gluon exchange, and turn to the survey of all six diagonal channels in Sec.~\ref{sec:diagonal}.

\subsection{The top channel and the strong coupling}
\label{sec:top}

\begin{figure}[t]
 \centering
 \includegraphics[width=\columnwidth]{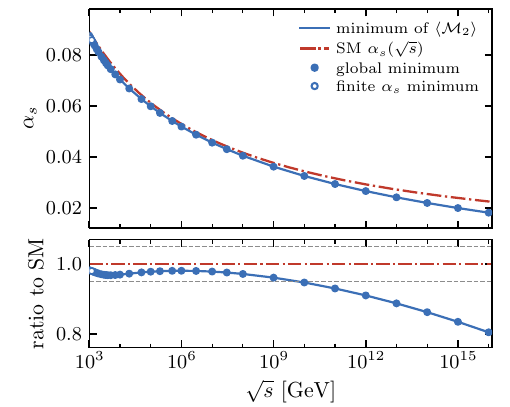}
 \caption{Magic-minimizing $\alpha_s$ in the $t\bar t$ color singlet as a function of $\sqrt{s}$, compared with the SM running coupling (dash-dotted). Filled circles: global minimum. Open circles:  finite $\alpha_s$ minimum, as the pure-QCD limit is lower. The lower panel shows the ratio to the SM value; the dashed lines mark $\pm5\%$.}
 \label{fig:alphas}
\end{figure}

Figure~\ref{fig:ttcurves} shows the magic in the $t\bar t$ singlet as a function of $\alpha_s=\alpha\,\rs$ at $\sqrt{s}=1\TeV$, $10\TeV$, and $10^8\GeV$. At each CM energy the SM  $\alpha_s(\sqrt{s})$ sits very close to the finite $\alpha_s$ minimum. At $10\TeV$ and $10^8\GeV$ the finite $\alpha_s$ minimum lies below the pure-QCD limit and is therefore the global minimum of Eq.~\eqref{eq:average}. At $1\TeV$ the finite $\alpha_s$ minimum is slightly above the pure-QCD limit, and the magic falls below it only for $\alpha_s>0.36$, a somewhat large coupling. The additional  structures at smaller $\alpha_s$ arise from interference among different Feynman diagrams. 

In Fig.~\ref{fig:alphas} we show the location of the finite $\alpha_s$ minimum from $1\TeV$ to $10^{16}\GeV$, together with the SM running coupling.  From $1.3\TeV$ it is the global minimum on the whole range $\rs\ge0$ (filled circles); between $1$ and $1.2\TeV$ the pure-QCD limit is lower (open circles), but the magic falls below the basin only for $\alpha_s$ above $0.36$ to $0.67$. Below $1\TeV$, where $2m_t/\sqrt{s}$ exceeds $0.3$, the structure changes: below $0.8\TeV$ a dip at $\alpha_s\approx0.03$ is deeper within $\alpha_s\le0.25$, and below $0.55\TeV$ there is no minimum near the SM point. Interestingly, the magic-minimizing $\alpha_s$ lies below the SM value at every CM energy above 1 TeV, as can be seen from Fig.~\ref{fig:alphas}. The relative deviation is $100(\alpha_s^{\min}/\alpha_s^{\SM}-1)$, which is only a few percent between  $\sqrt{s}=1$ TeV and $10^8$ GeV.

\subsection{The importance of the Higgs}
\label{sec:mechanism}

\begin{figure}[t]
 \centering
 \includegraphics[width=\columnwidth]{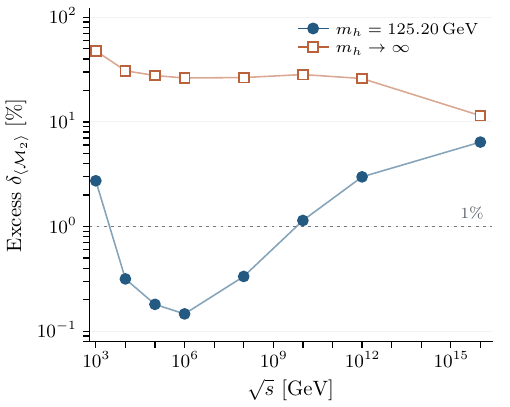}
 \caption{Percentage excess $\delta_{\Mtwo}$ in the $t\bar t$ color singlet as a function of the CM energy for $m_h=125.20\GeV$ (filled circles) and $m_h\to\infty$ (open squares). Each curve uses its own minimum over $\alpha_s$, with all other inputs held fixed at each CM energy.}
 \label{fig:higgs}
\end{figure}

\begin{figure*}[t]
 \centering
 \includegraphics[width=0.9\textwidth]{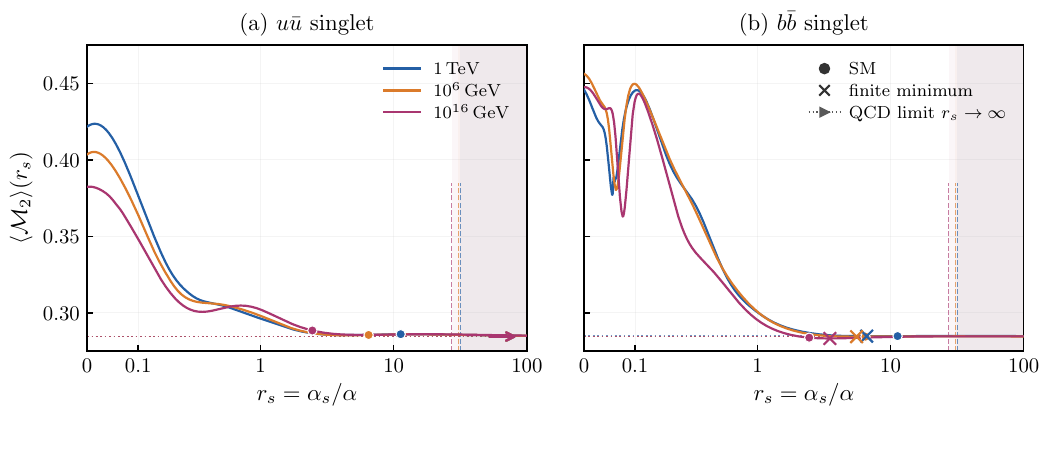}
 \caption{Magic in the $u\bar u$ (left) and $b\bar b$ (right) color singlets at $\sqrt{s}=1\TeV$, $10^6\GeV$, and $10^{16}\GeV$. Curves extend to $\rs=100$; color-matched shading marks $\alpha_s>0.25$. Circles denote SM ratios and crosses finite minima; arrows identify asymptotic minima on the QCD reference lines. The $c\bar c$ curves differ from $u\bar u$ by at most $5\times10^{-5}$, and the $d\bar d$ and $s\bar s$ curves from $b\bar b$ by at most $2\times10^{-4}$ for $\rs>1$. }
 \label{fig:lightcurves}
\end{figure*}

\begin{figure*}[t]
 \centering
 \includegraphics[width=0.9\textwidth]{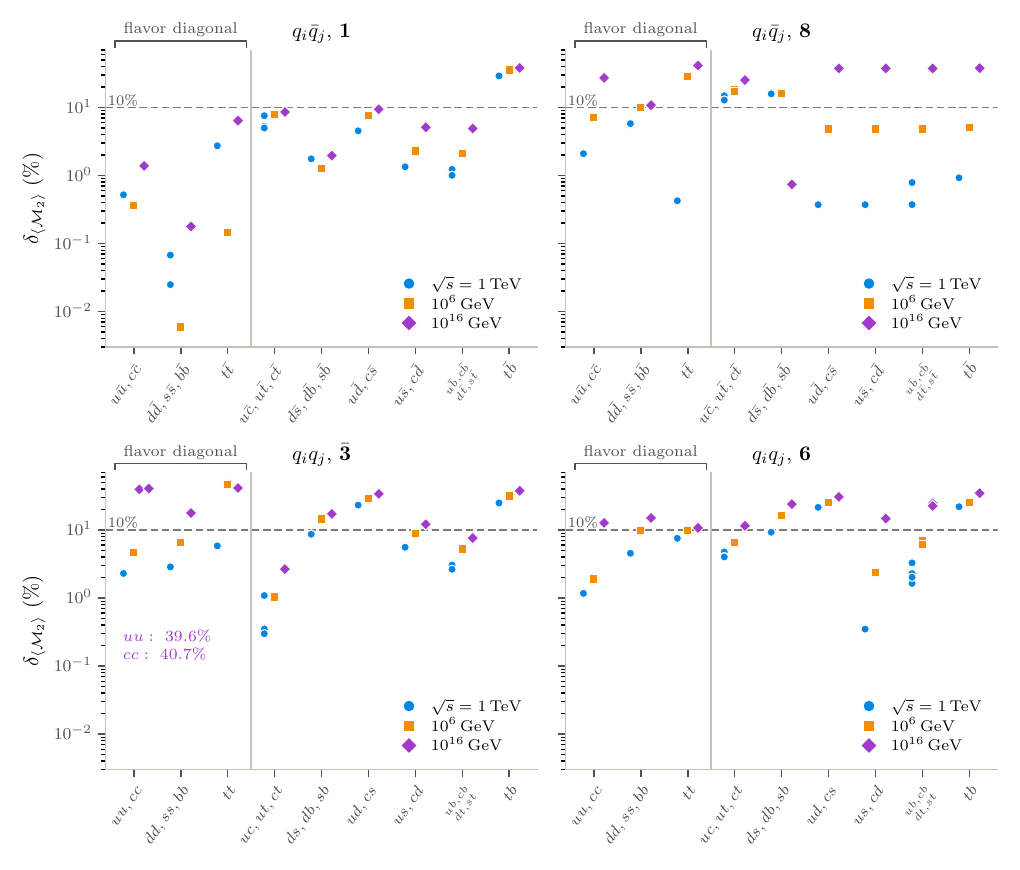}
 \caption{$\delta_{\Mtwo}$ in all  quark channels at $\sqrt{s}=1\TeV$, $10^6\GeV$, and $10^{16}\GeV$: $q_i\bar q_j$ in the color singlet and octet (top) and $q_iq_j$ in the antitriplet and sextet (bottom), for all 21 flavor pairs.  Channels with the same electroweak charges and, for up-down pairs, comparable $|V_{ij}|^2$ are grouped on the horizontal axis; the $u\bar b,c\bar b$ position also contains $t\bar d$ and $t\bar s$. The first three positions contain the flavor-diagonal channels, separated from the remaining channels by the vertical line. Blue circles, orange squares, and purple diamonds denote the three CM energies, respectively. Small horizontal offsets distinguish the $uu$ (left) and $cc$ (right) antitriplet points at $10^{16}\GeV$. The dashed horizontal line marks $\delta_{\Mtwo}=10\%$.}
 \label{fig:allchannels}
\end{figure*}

Higgs exchange is essential for this result. At $1\TeV$ and $m_h=125.20\GeV$, the magic at the SM value of $\alpha_s$ is $0.298384$, close to the global minimum $0.290433$ at the pure-QCD limit. In the limit $m_h\to\infty$, the magic at the same value of $\alpha_s$ rises to $0.335687$, while the minimum is $0.227095$ at $\alpha_s=0$. The corresponding excesses are
\begin{equation}
 \begin{aligned}
 m_h=125.20\GeV:&\quad \delta_{\Mtwo}=2.7\%,\\
 m_h\to\infty:&\quad \delta_{\Mtwo}=47.8\%.
 \end{aligned}
 \label{eq:top}
\end{equation}
Removing the Higgs here means taking the $m_h\to\infty$ limit at each fixed finite CM energy, with all other parameters held fixed. This limit removes both the $t$- and $s$-channel physical-Higgs exchange diagrams, while leaving all other contributions unchanged. The $t$- and $s$-channel diagrams contribute with opposite sign: removing only the $s$-channel Higgs exchange raises the SM-point magic to $0.348331$, whereas removing only the $t$-channel lowers it to $0.288710$. The SM result reflects both contributions together.

Figure~\ref{fig:higgs} extends the comparison to higher CM energies. At the physical Higgs mass, $\delta_{\Mtwo}$ stays below $3\%$ at all sampled CM energies from $1\TeV$ to $10^{12}\GeV$. Taking the $m_h\to\infty$ limit, it lies between $26\%$ and $48\%$ up to $10^{12}\GeV$,  with the magic minimum always sitting at $\alpha_s=0$; at $10^{16}\GeV$ it is $11.5\%$, measured from the pure-QCD limit.

\subsection{Light flavor-diagonal singlets}
\label{sec:diagonal}

In this subsection we consider the remaining five flavor-diagonal color singlets.  Fig.~\ref{fig:lightcurves}  shows the magic production as a function of $\rs=\alpha_s/\alpha$  for the $u\bar u$ and $b\bar b$ singlets, representing the up-type and down-type quarks.  The $c\bar c$ curves are almost identical to $u\bar u$, differing  by at most $5\times10^{-5}$ in $\Mtwo$,  while the $d\bar d$ and $s\bar s$ curves differ from $b\bar b$  by at most $2\times10^{-4}$ for $\rs>1$ and $2.4\times10^{-3}$ at smaller $\rs$, except near the sharp feature at $\rs\approx0.05$ to $0.09$, where the difference is approximately $1.2\times10^{-2}$.

The two panels show two qualitatively different behaviors. The $b\bar b$ singlet, the weak-isospin partner of the top, descends from a sharp feature at small $\rs$ to a shallow finite minimum slightly below the pure-QCD value at all three CM energies; the SM magic exceeds these minima by at most $0.18\%$. The up-type channels approach the pure-QCD value from above, and their shallow local dips lie above it, so that their minimum is the pure-QCD limit. 

At CM energies from $1\TeV$ to $10^{12}\GeV$, the five lighter flavors lie within $1\%$ of their minima. At $10^{16}\GeV$ the $u\bar u$ and $c\bar c$ excesses reach $1.38\%$, while the three down-type channels remain below $0.18\%$. For these five flavors, physical Higgs exchange changes the excess by at most $0.0012$ percentage points, in contrast to the top channel of Sec.~\ref{sec:mechanism}.

\begin{table}[t]
\caption{Increase of the magic when $\alpha_s$ is lowered from its SM value to $\alpha$, $100\,[\Mtwo(\rs=1)/\Mtwo(\rs^{\SM})-1]$, in the six diagonal color singlets, evaluated directly at $\rs=\alpha_s/\alpha=1$ and $\rs^{\SM}$.}
\label{tab:weak}
\centering\setlength{\tabcolsep}{7pt}
\begin{tabular}{lccc}
\toprule
channel & $1\TeV$ & $10^6\GeV$ & $10^{16}\GeV$\\
\midrule
$u\bar u$ & $3.57\%$ & $4.37\%$ & $4.74\%$\\
$c\bar c$ & $3.57\%$ & $4.37\%$ & $4.74\%$\\
$d\bar d$ & $5.42\%$ & $5.69\%$ & $4.05\%$\\
$s\bar s$ & $5.42\%$ & $5.69\%$ & $4.05\%$\\
$b\bar b$ & $5.46\%$ & $5.71\%$ & $4.05\%$\\
$t\bar t$ & $14.45\%$ & $38.04\%$ & $16.54\%$\\
\midrule
$\rs^{\SM}$ & $11.29$ & $6.49$ & $2.45$\\
\bottomrule
\end{tabular}
\end{table}

While the minimization of magic in the five light-flavor channels does not allow for a precise determination of the SM $\alpha_s$, they might provide clues as to why the strong coupling constant is ``strong'' from the quantum-computational viewpoint.  In all five channels the magic is largest when the strong coupling is weak. In Fig.~\ref{fig:lightcurves} it lies between $0.38$ and $0.46$ at $\alpha_s=0$, falls to about $0.30$ at $\alpha_s=\alpha$, and comes within $1\%$ of the pure-QCD value once $\alpha_s$ exceeds about $3\alpha$. The SM coupling, with $\alpha_s/\alpha=11.3$ at $1\TeV$ and $2.5$ at $10^{16}\GeV$, lies in or near this low-magic region. Table~\ref{tab:weak} quantifies the comparison with a coupling of electromagnetic strength: lowering $\alpha_s$ from its SM value to $\alpha$ increases the magic by $3.6\%$ to $5.7\%$ in the five lighter flavors, and by $14\%$ to $38\%$ in the top channel, at the three CM energies of Fig.~\ref{fig:lightcurves}. Within the low-magic region, however, the curves are too shallow to single out a value of $\alpha_s$.

\subsection{Flavor-nondiagonal and nonsinglet channels}
\label{sec:other}

Figure~\ref{fig:allchannels} summarizes the excesses in all flavor and color channels at $1\TeV$, $10^6\GeV$, and $10^{16}\GeV$. The light flavor-nondiagonal singlets lie $1.2\%$ to $9.4\%$ above their minima, somewhat further than the diagonal ones. Among the light-quark nonsinglets, the excess reaches $40.7\%$ in the $cc$ antitriplet at $10^{16}\GeV$, while the $uu$ antitriplet gives $39.6\%$. There is no uniform ordering with color: the $u\bar d$ octet has $0.37\%$ at $1\TeV$, $4.84\%$ at $10^6\GeV$, and $37.6\%$ at $10^{16}\GeV$, whereas the $d\bar s$ octet has $15.9\%$ at $1\TeV$ and at $10^6\GeV$ but $0.74\%$ at $10^{16}\GeV$. The $t\bar t$ octet gives $28.7\%$ at $10^6\GeV$ and $41.4\%$ at $10^{16}\GeV$.

In the massive color-singlet $t\bar b$ channel, which also admits charged-current exchange, the excess is $29.2\%$ at $1\TeV$, $35.7\%$ at $10^6\GeV$, and $38.2\%$ at $10^{16}\GeV$. At $1\TeV$, removing physical Higgs exchange changes the excess, with the minimum recomputed for the modified amplitude, by only $0.034$ percentage points, whereas removing charged-current exchange reduces it to $1.00\%$. This comparison indicates that the large excess is associated with charged-current exchange rather than simply with the presence of the top quark. 

Finally, the preference for $\alpha_s>\alpha$ found in the flavor-diagonal color-singlet channels does not extend to these channels. At $\alpha_s=\alpha$ the magic is lower than at the SM point in the $u\bar c$ singlet at every sampled CM energy from $\mu_0$ to $10^{16}\GeV$, and at some CM energies also in the $u\bar c$ octet and the $d\bar s$ singlet (at  $10^{14}$ and $10^{16}\GeV$) and in the $uu$ sextet and antitriplet (at $10^{14}$ and $10^{16}\GeV$). In the $ud$ antitriplet even the pure electroweak amplitude, $\alpha_s=0$, produces $1\%$ to $4\%$ less magic than the SM point at every sampled CM energy from $1\TeV$ to $10^{16}\GeV$.

\section{Discussion}
\label{sec:discussion}

In this work we have studied quantum magic in elastic quark scattering at tree level and obtained two complementary results. In the color-singlet $t\bar t$ channel, varying only $\alpha_s$, with all other inputs held fixed at each CM energy, yields a finite-$\alpha_s$ magic minimum that reproduces the running $\overline{\mathrm{MS}}$ strong coupling to within a few percent from approximately $1\TeV$ to $10^8\GeV$. This minimum becomes global above approximately $1.3\TeV$, and physical Higgs exchange is essential for the agreement. The five lighter flavor-diagonal singlets provide a complementary result: although their minima are too shallow to determine $\alpha_s$ accurately, they favor a strong interaction stronger than electromagnetism. Reducing $\alpha_s$ from its SM value to the electromagnetic coupling increases magic production in all six channels. Together with the earlier determination of the weak mixing angle from charged-lepton scattering~\cite{Liu:2025bgw}, these findings suggest that magic minimization may account for both the values of interaction parameters and their relative strengths. 

These findings are mysterious. The gauge couplings are fundamental constants because their values are determined experimentally rather than predicted within the theory, and they enter calculations as input parameters of the SM. Why, then, should magic minimization reproduce both the weak mixing angle and the strong coupling with such precision?

A radical possibility is that the parameters of fundamental interactions among color singlets reflect an underlying principle of {\bf quantum computational efficiency}. Magic provides a resource for universal quantum computation beyond Clifford operations, giving a concrete computational meaning to the quantity being minimized~\cite{Bravyi:2004isx}. This motivates the conjecture that the physical universe emerges from a quantum simulation subject to constrained resources, with its interaction parameters selected by quantum resource minimization.

Indeed, a physical process can itself be viewed as a computation: the initial state supplies the input, the final state provides the output, and the dynamics specifies the algorithm. From this perspective, our findings suggest that SM dynamics may favor algorithms that minimize the quantum resources required to carry out these computations. Such a principle could account for the quantitative agreement found for both the weak mixing angle and the strong coupling, as well as the qualitative preference for a strong interaction stronger than electromagnetism.

Several developments provide scientific motivation for exploring this possibility. The dynamics of quantum systems with local interactions can be simulated efficiently on a quantum computer, as envisioned by Feynman~\cite{Feynman:1982} and established in Ref.~\cite{Lloyd:1996aai}. For example, in scalar quantum field theories, an explicit quantum algorithm for relativistic scattering was constructed in Ref.~\cite{Jordan:2012xnu}. Physical bounds on information storage and processing give resource constraints a concrete meaning within physical law~\cite{Lloyd:2001bh}.  Possible observable consequences of a classical spacetime-lattice simulation have also been investigated~\cite{Beane:2012rz}, while the quantum error correction interpretation of bulk locality in AdS/CFT connects spacetime properties to quantum information~\cite{Almheiri:2014lwa}. These studies motivate investigating a computational origin of physical laws. 

Our conjecture concerns a complementary possibility: that constraints on quantum computational resources determine the values of the interaction parameters themselves. Our findings in Ref.~\cite{Liu:2025bgw} and the present work suggest several requirements for such a principle and identify concrete questions for further study:

\begin{itemize}
\item \textbf{Which measure of magic is physically relevant?} A resource principle must specify the quantity being minimized. Different measures of magic need not select the same values of the interaction parameters, allowing comparison with experiment to discriminate among candidate measures. If minimizing the SRE consistently reproduces the observed parameters across independent processes while other measures do not, this would favor the SRE as a quantity governing the selection of fundamental constants. The second SRE used here also has practical advantages: for the pure two-qubit states considered, it is precisely defined and straightforward to evaluate from Pauli expectation values, without an optimization over stabilizer decompositions~\cite{Leone:2021rzd}. The agreements found for the weak mixing angle and the strong coupling motivate systematic comparisons with other measures, making the choice of resource measure itself an empirically testable part of the conjecture.

\item \textbf{Which processes enter the minimization?} The process dependence observed here raises the question of how different processes contribute to a common resource cost. Color neutrality alone does not ensure near-minimal magic, as illustrated by the $t\bar b$ singlet, whose large excess is associated with charged-current exchange. Since this amplitude is proportional to $\lvert V_{tb}\rvert^2$, the channel may incur a resource cost associated with parameters favored by other processes. CKM unitarity ties a reduction of $\lvert V_{tb}\rvert^2$ to an increase in $\lvert V_{td}\rvert^2+\lvert V_{ts}\rvert^2$, suggesting that several channels may need to be considered together. The conjecture directs our attention to physical color-singlet states. Applying it fully requires understanding the relation between the perturbative quark singlets considered here and physical hadronic processes.

\item \textbf{What is the computational prescription?} The Lab basis provides a common encoding for incoming and outgoing spins along the fixed beam axis. Expressing the outgoing spins in the helicity basis requires angle-dependent rotations, which are generally non-Clifford and can change the magic~\cite{Li:2026zum}. Whether the Lab encoding minimizes the total computational cost depends on the implementation of the scattering process. Moreover, the present analysis concerns spin magic with the external flavors specified. Extending the framework to other quantum numbers, such as flavor, and treating spin and flavor jointly would allow a broader assessment of the resources required by the dynamics.

\item \textbf{Which SM parameters can the principle explain?} The minimal SM with massless neutrinos contains nineteen independent parameters. A resource principle need not determine all of these parameters. Understanding which it can explain and which it leaves undetermined may be equally informative, helping to establish its scope and regime of validity.\footnote{Ref.~\cite{Li:2026udy} found a local minimum of averaged spin magic at the strong CP phase $\bar\theta=0$ for certain ranges of the CP-conserving spin interaction in low-energy neutron-proton scattering. Whether this preference persists in a complete hadronic scattering calculation remains to be established.} Other processes could provide independent criteria for determining the Yukawa couplings, flavor mixing angles, CP-violating phases, and Higgs-sector parameters that are used as fixed inputs here. Going outside of spin degrees of freedom, studying magic in flavor degrees of freedom could be particularly useful for investigating the flavor structure of these interactions.

\item \textbf{Can the principle make quantitative predictions?} Correlated scans would test whether parameter values preferred by different processes are mutually compatible and whether additional minima or flat directions emerge when several parameters vary together. Radiative corrections are needed to establish the stability of the minima and the theoretical precision of the inferred parameters. A particularly compelling test would be a prediction for a parameter whose value is not yet well determined experimentally, with controlled uncertainties that permit an independent experimental test.

\item \textbf{What role does the Higgs sector play?} The essential role of Higgs exchange motivates asking whether resource minimization could constrain the Higgs sector itself. At fixed finite CM energy, taking $m_h\to\infty$ with all other inputs to the tree amplitudes held fixed is equivalent to removing both the $s$- and $t$-channel physical-Higgs exchange diagrams by hand. In our fixed-$v$ prescription, the tree-level relation $m_h^2=2\lambda v^2$ identifies this as the formal $\lambda\to\infty$ limit of the Higgs quartic coupling. At $\sqrt{s}=1\TeV$, this increases the magic at the SM coupling and moves it much further from its minimum. Could the same principle favor a finite Higgs mass and shed light on its small value relative to higher physical scales? The present comparison does not determine the preferred mass. Addressing this question requires a consistent treatment of the Higgs potential and a demonstration that any preferred mass remains stable under radiative corrections.
\end{itemize}

Such investigations could reveal whether quantum computational efficiency provides a predictive principle for fundamental interactions.

Finally, in the separate scans of $\sin\theta_w$ and $\alpha_s$, with all other inputs held fixed, we find no coupling values for which magic production vanishes.  Nonstabilizer states, which have non-zero magic, together with Clifford operations and Pauli measurements, enable universal quantum computation~\cite{Bravyi:2004isx,Reichardt:2004magic}. Our results therefore motivate considering a hypothetical universe restricted to stabilizer states, Clifford gates, and Pauli measurements. Such a universe could still exhibit superposition and entanglement, while admitting efficient classical simulation and lacking universal quantum computational capability. How would the laws of physics in such a universe differ from those of our own? We leave this question for future exploration.

\section*{acknowledgments}
We thank Jesse Thaler and Carlos Wagner for feedbacks on the manuscript. We  acknowledge the use of Claude (Anthropic) and ChatGPT (OpenAI) for assistance with computations, writing, and figure generation. The authors take full responsibility for the correctness and scientific value of the work. Z.Y. is supported
by IHEP under Grant No. E5515AU1. I.L. is supported in part by the U.S. Department of Energy under contracts DE-AC02-06CH11357 (Argonne), DE-SC0023522 (Northwestern), DE-SC0010143 (Northwestern) and No.~89243024CSC000002 (QuantISED Program).

\interlinepenalty=10000
\bibliography{literature}
\end{document}